\documentstyle[amssymb,preprint,aps]{revtex}

\begin{document}
\title{Electric-field-induced nematic-cholesteric transition and 3-D
director structures in homeotropic cells}
\author{I. I. Smalyukh ($\thanks{%
Author for correspondence (e-mail: smalyukh@lci.kent.edu)}$), B. I. Senyuk,
P. Palffy-Muhoray, and O. D. Lavrentovich}
\address{Liquid Crystal Institute and Chemical Physics \\
Interdisciplinary Program, Kent State University, Kent, Ohio 44242}
\author{H. Huang\ and\ E. C. Gartland, Jr.}
\address{Department of Mathematical Sciences, Kent State University, Kent,\\
Ohio 44242}
\author{V. H. Bodnar, T. Kosa, and B. Taheri}
\address{AlphaMicron Inc., Kent, Ohio 44240.}
\date{\today}
\maketitle

\begin{abstract}
We study the phase diagram of director structures in cholesteric liquid
crystals of negative dielectric anisotropy in homeotropic cells of thickness 
$d$ which is smaller than the cholesteric pitch $p$. The basic control
parameters are the frustration ratio $d/p$ and the applied voltage $U$. Upon
increasing $U$, the direct transition from completely unwound homeotropic
structure to the translationally invariant configuration ($TIC$) with
uniform in-plane twist is observed at small $d/p\lessapprox 0.5$.
Cholesteric fingers that can be either isolated or arranged periodically
occur at $0.5\lessapprox d/p<1$ and at the intermediate $U$ between the
homeotropic unwound and $TIC$ structures. The phase boundaries are also
shifted by (1) rubbing of homeotropic substrates that produces small
deviations from the vertical alignment; (2) particles that become nucleation
centers for cholesteric fingers; (3) voltage driving schemes. A novel
re-entrant behavior of $TIC$ is observed in the rubbed cells with
frustration ratios $0.6\lessapprox d/p\lessapprox 0.75,$ which disappears
with adding nucleation sites or using modulated voltages. In addition,
Fluorescence Confocal Polarising Microscopy (FCPM) allows us to directly and
unambiguously determine the 3-D director structures. For the cells with
strictly vertical alignment, FCPM confirms the director models of the
vertical cross-sections of four types of fingers previously either obtained
by computer simulations or proposed using symmetry considerations. For
rubbed homeotropic substrates, only two types of fingers are observed, which
tend to align along the rubbing direction. Finally, the new means of control
are of importance for potential applications of the cholesteric structures,
such as switchable gratings based on periodically arranged fingers and
eyewear with tunable transparency based on $TIC$.
\end{abstract}

\section{Introduction}

The unique electro-optic and photonic properties of cholesteric liquid
crystals (CLCs) make them attractive for applications in displays,
switchable diffraction gratings, eyeglasses with voltage-controlled
transparency, for temperature visualization, for mirrorless lasing, in beam
steering and beam shaping devices, and many others \cite%
{Blinov-Ch-book,McManamon,NatureChColor,DKYang,EyeLC1,Patel,ChGrating1,Bistable-chiral,ChLaser1,ChLaserAPL,ChLaser3,de Gennes-book,KlemanLavrentovichBook,ChiralityInLC}%
. In nearly all of these applications, CLCs are confined between flat glass
substrates treated to set the orientation of molecules at the liquid crystal
(LC)-glass interface along some well-defined direction (called easy axis)
and an electric field is often used to switch between different textures. In
the confined CLCs, the magnitude of the free energy terms associated with
elasticity, surface anchoring, and coupling to the applied field are
frequently comparable; their competition results in a rich variety of
director structures that can be obtained by appropriate surface treatment,
material properties of CLCs, and applied voltage. Understanding these
structures and transitions between them is of great practical interest and
of fundamental importance\cite{de
Gennes-book,KlemanLavrentovichBook,ChiralityInLC}.

CLCs have a twisted helicoidal director field in\ the ground state. The axis
of molecular twist is called the helical axis and the spatial period over
which the liquid crystal molecules twist through $2\pi $ is called the
cholesteric pitch $p$. CLCs can be composed of a single compound or of
mixtures of a nematic host and one or more chiral additives. Cholesterics
usually have the equilibrium pitch $p$ in the range $100nm-100\mu m$; the
pitch $p$ can be easily modified by additives. When CLCs are confined in the
cells with different boundary conditions or subjected to electric or
magnetic fields, one often observes complex three-dimensional (3-D)
structures. The cholesteric helix can be distorted or even completely
unwound by confining CLCs between two substrates treated to produce
homeotropic boundary conditions \cite{Oswald-review00}. Interest in this
subject was initiated by Cladis and Kleman \cite{CladisKleman}, subsequently
a rich variety of spatially periodic and uniform structures have been
reported \cite%
{Oswald-review00,LGil,ourreview,PressArrott,MCLC-PressArrott,Flow-PressArrott,Gil-PRL98,Baudry-PRE99,CHnegativeDeltaE,Tarasov,Oswald2004,InvWalls,T-Junktions,FLequeux,Ishikawa-fingerprint}%
. \ These structures can be controlled by varying the cell gap thickness $d$%
, pitch $p$, applied voltage $U$, and the dielectric and elastic properties
of the used CLC. The complexity of many LC structures usually does not allow
simple analytic descriptions of the director configuration. Since the
pioneering work of Press and Arrott\cite{PressArrott,MCLC-PressArrott}, a
great progress has been made in computer simulations of static director
patterns in CLCs confined into homeotropic cells (see, for example \cite%
{Oswald-review00,LGil,ourreview,PressArrott,Flow-PressArrott,Gil-PRL98,Baudry-PRE99}%
), which brought much of the current understanding of these structures.

The first goal of this work is to study phase diagram and director
structures that appear because of geometrical frustration of CLCs in the
cells with either strictly vertical or slightly tilted ($<2^{\circ }$) easy
axis at the confining substrates. We start with the phase diagram in the
plane of $\rho =d/p$ and $U$ similar to the one reported in Refs.\cite%
{Oswald-review00,CHnegativeDeltaE} and then proceed by studying influence of
such extra parameters as rubbing, introducing nucleation sites, and voltage
driving schemes. We use CLCs with negative dielectric anisotropy; the
applied voltages are sufficiently low and the frequencies are sufficiently
high to avoid hydrodynamic instabilities \cite{de Gennes-book}. Cell gap
thicknesses $d$ are smaller than $p$ and the phase diagrams are explored for
frustration ratios $\rho =0-1$ and $U=(0-4)Vrms$. For small $\rho $ and $U$,
the boundary conditions force the LC molecules throughout the sample to
orient perpendicular to the glass plates. Above the critical values of $\rho 
$ and/or $U$, cholesteric twisting of the director takes place \cite%
{Oswald-review00}. Depending on $\rho ,$ $U,$ and other conditions, the
twisted director structures can be either uniform or spatially periodic,
with wave vector in the plane of the cell. Upon increasing $U$ for $\rho
<0.5 $, the direct transition from completely unwound homeotropic structure
to the translationally invariant configuration ($TIC$) \cite%
{Oswald-review00,PressArrott,MCLC-PressArrott} with uniform in-plane twist
is observed. Cholesteric fingers ($CF$s) of different types that can be
either isolated or arranged periodically are observed for $0.5\lessapprox
\rho <1$ and intermediate $U$ between the homeotropic unwound and $TIC$
structures. The phase diagrams change if the homeotropic alignment layers
are rubbed, if particles that become the nucleation centers for $CF$s are
present, and if different driving voltage schemes are used. Upon increasing
U in rubbed homeotropic cells with $0.6<\rho <0.75$, we observe a re-entrant
behavior of $TIC$ and the following transition sequence: (1) homeotropic
untwisted state, (2) translationally invariant twisted state, (3) periodic
fingers structure (4) translationally invariant twisted state with larger
in-plane twist. This sequence has not been observed in our own and in the
previously reported\cite{Oswald-review00} studies of unrubbed homeotropic
cells.

The second goal of this work is to unambiguously reconstruct director field
of $CF$s and other observed structures in the phase diagram. For this we use
the Fluorescence Confocal Polarizing Microscopy (FCPM) \cite{FCPMCPL}.
Although the fingers of different types look very similar under the
polarizing microscope (which may explain some confusion in the literature %
\cite{Baudry-PRE99}), FCPM allows clear differentiation of $CF$s, as well as
other structures. We directly visualize the $TIC$ with the total director
twist ranging from $0$ to $2\pi $, depending on $\rho $ and $U,$ and
rubbing. We reconstruct director structure in the vertical cross-sections of
four different types of $CF$s that are observed for cells with strictly
vertical alignment. We unambiguously prove the models described recently by
Oswald et al. \cite{Oswald-review00} while disproving some of the other
models that were proposed in the early literature (see, for example\cite%
{Baudry-PRE99}). Only two types of $CF$ structures are observed in CLCs
confined to cells with slightly tilted easy axes at the substrates.

The third goal of our work stems from the importance of the studied
structures for practical applications. The spatially-uniform $TIC$ and
homeotropic-to-$TIC$ transition are used in the electrically driven light
shutters, intensity modulators, eyewear with tunable transparency, and
displays \cite{Blinov-Ch-book,EyeLC1,Patel,Bistable-chiral}. In these
applications, it is often advantageous to work in the regime of high $\rho $
but fingers are not desirable since they scatter light. In our study, we
therefore focus on obtaining maximum effect of different factors on the
phase diagrams. We demonstrate that the combination of rubbing and
low-frequency voltage modulation can stabilize the uniformly twisted
structures up to $\rho \approx 0.75$, much larger than $\rho \approx 0.5$
reported previously \cite{Oswald-review00}. The presence of nucleation
centers, such as particles used to set the cell thickness, tends to destroy
homogeneously twisted cholesteric structure even at relatively low $\rho
\approx 0.5$ confinement ratios; this information is important for the
optimal design of the finger-free devices. On the other hand, periodic
finger patterns with well controlled periodicity and orientation may be used
as voltage-switchable diffraction gratings. Our finding, which enables the
very possibility of such application, is that rubbing can set the
unidirectional orientation of periodically arranged fingers.

The article is organized as follows. We describe materials, cell
preparation, and experimental techniques in section II. The phase diagrams
are described in section III.A and the reconstructed director structures in
section III.B. Section IV gives an analytical description of the transition
from homeotropic to a twisted state as well as a brief discussion of other
structures and transitions along with their potential applications. The
conclusions are drawn in section V.

\section{Experiment}

\subsection{Materials and cell preparation}

The cells with homeotropic boundary conditions were assembled using glass
plates coated with transparent ITO electrodes and the polyimide JALS-204
(purchased from JSR, Japan) as an alignment layer. JALS-204 provides strong
homeotropic anchoring; anchoring extrapolation length, defined as the ratio
of the elastic constant to the anchoring strength, is estimated to be in the
submicron range. Some of the substrates with thin layers of JALS-204 were
unidirectionally buffed (5 times using a piece of velvet cloth) in order to
produce an easy axis at a small angle $\gamma $ to the normal to the cell
substrates. $\gamma $ was measured by conoscopy and magnetic null methods %
\cite{MagneticNull}. \ The value of $\gamma $ weakly depends on the rubbing
strength, but in all cases it was small, $\gamma <2^{\circ }$. The cell gap
thickness was set using either the glass micro-sphere spacers uniformly
distributed within the area of a cell (one spacer per approximately $100\mu
m\times 100\mu m$ area) or strips of mylar film placed along the cell edges.
The cell gap thickness $d$ was measured after cell assembly using the
interference method \cite{BornWolf} with a LAMBDA18 (Perkin Elmer)
spectrophotometer. In order to study textures as a function of the
confinement ratio $\rho =d/p$, we constructed a series of cells, with
identical thickness, but filled with CLCs of different pitch $p$. To
minimize spherical aberrations in the FCPM, observations were made with
immersion oil objectives, using glass substrates of thickness $0.15mm$ with
refractive index $1.52$ \cite{FCPMCPL}. \ Regular ($1mm$) and thick ($3mm$)
substrates were used to construct cells for polarizing microscopy (PM)
observations.

Cholesteric mixtures were prepared using the nematic host AMLC-0010
(obtained from AlphaMicron Inc., Kent, OH) and the chiral additive ZLI-811
(purchased from EM Industries). The helical twisting power $HTP=10.47\mu
m^{-1}$ of the additive ZLI-811 in the AMLC-0010 nematic host was determined
using the method of Grandjean-Cano wedge \cite{SmalyukPRE,TKosaMCLC}. \ The
obtained mixtures had pitch $2<p<500\mu m$ as calculated from $%
p=1/(c_{chiral}\cdot HTP)$ where $c_{chiral}$ is the weight concentration of
the chiral agent, and verified by the Grandjean-Cano wedge method \cite%
{SmalyukPRE,TKosaMCLC,SmalyukhPRL}.\ The low frequency dielectric anisotropy
of the AMLC-0010\ host$\ $is $\Delta \varepsilon =-3.7$ ($\varepsilon
_{\parallel }=3.4,$ $\varepsilon _{\perp }=7.1$) as determined from
capacitance measurements for homeotropic and planar cells using an SI-1260
Impedance/Gain-phase analyzer (Schlumberger) \cite{Blinov-Ch-book,deJeu}.
The birefringence of AMLC-0010 is $\Delta n=0.078$ as measured with an Abbe
refractometer. \ The elastic constants describing the splay, twist, and bend
deformations of the director in AMLC-0010 are $K_{11}=17.2pN$, $%
K_{22}=7.51pN $, $K_{33}=17.9pN$ as determined from the thresholds of
electric and magnetic Freedericksz transition in different geometries \cite%
{Blinov-Ch-book,deJeu}. \ The cholesteric mixtures were doped with a small
amount of fluorescent dye
n,n'-bis(2,5-di-tert-butylphenyl)-3,4,9,10-perylenedicarboximide (BTBP) \cite%
{FCPMCPL} for the FCPM studies. Small quantities ( $0.01$wt. \% ) of BTBP
dye were added to the samples; at these concentrations, the dye is not
expected to affect properties of the CLCs used in our studies.

Constant amplitude and modulated amplitude signals were applied to the cells
using a DS345 generator (Stanford Research Systems) and a Model 7602
Wide-band Amplifier (Krohn-Hite) which made possible the use of a wide range
of carrier and modulation frequencies ($10-100000$) $Hz$. The transitions
from the homeotropic untwisted to a variety of twisted structures were
monitored via capacitance measurements and by measuring the light
transmittance of the cell between crossed polarizers. The transitions
between different director structures and textures were characterized with
PM and FCPM \cite{FCPMCPL} as described below.

\subsection{Polarizing Microscopy and Fluorescence Confocal Polarizing
Microscopy}

Polarizing Microscopy (PM) observations were performed using the Nikon
Eclipse E600 POL microscope with the Hitachi HV-C20 CCD camera. The PM\
studies were also performed using a BX-50 Olympus microscope in the PM mode.
In order to directly reconstruct the vertical cross-sections of the
cholesteric structures, we performed further studies in the FCPM mode of the
very same modified BX-50 microscope\cite{FCPMCPL} as described below. The PM
and FCPM techniques are used in parallel and provide complementary
information.

The FCPM set-up was based on a modified BX-50 fluorescence confocal
microscope \cite{FCPMCPL}. The excitation beam ($\lambda =488nm$, from an Ar
laser) is focused by an objective onto a small submicron volume within the
CLC cell. The fluorescent light from this volume is detected by a
photomultiplier in the spectral region $510-550nm$. A pinhole is used to
discriminate against emission from the regions above and below the selected
volume. The pinhole diameter $D$ is adjusted according to magnification and
numerical aperture ${\rm NA}$ of the objective; $D=$ $100\,\mu m$ for an
immersion oil $60\times $ objective with ${\rm NA}=1.4$. A very same
polarizer is used to determine the polarization of both the excitation beam
and the detected fluorescent light collected in the epifluorescence mode.
The relatively {low birefringence (}${\Delta n\approx 0.0}78$) {of the
AMLC-0010 nematic host }mitigates two problems that one encounters in FCPM
imaging of CLCs: (1) defocussing of the extraordinary modes relative to the
ordinary modes \cite{FCPMCPL} and (2) the Mauguin effect, where polarization
follows the twisting director field \cite{ourreview,Yehbook}.

The used BTBP dye has both absorption and emission transition dipoles
parallel to the long axis of the molecule \cite%
{SmalyukPRE,FordKamat,SmalyukhPRL}. The FCPM signal, resulting from a
sequence of absorption and emission events, strongly depends on the angle $%
\beta $ between the transition dipole moment of the dye (assumed to be
parallel to the local director ${\bf \hat{n}}$) and the polarization ${\bf 
\hat{P}}$.$\ $The intensity scales as $I_{FCPM}\sim $ $\cos ^{4}\beta $, %
\cite{FCPMCPL} as both the absorption and emission are proportional to $\cos
^{2}\beta ${\bf .} The strongest FCPM signal corresponds to ${\bf \hat{n}%
\parallel \hat{P}}$, where $\beta =0$, and sharply decreases when $\beta $
becomes non-zero \cite{ourreview,FCPMCPL,SmalyukPRE,SmalyukhPRL}. By
obtaining the FCPM images for different ${\bf \hat{P}}$, we reconstruct
director structures in both in-plane and vertical cross-sections of the cell
from which then the entire 3-D director pattern is reconstructed. We note
that in the FCPM images the registered fluorescence signal from the bottom
of the cell can be somewhat weaker than from the top, as a result of light
absorption, light scattering caused by director fluctuations,
depolarization, and defocussing. To mitigate these experimental artefacts
and to maintain both axial and radial resolution within $1\mu m$, we used
relatively shallow ($\leq $ $20$ $\mu m$) scanning depths \cite%
{ourreview,FCPMCPL}. The other artefacts, such as light depolarization by a
high NA objective, are neglected as they are of minor importance \cite%
{ourreview,FCPMCPL,SmalyukPRE,SmalyukhPRL}.

\section{Results}

\subsection{Phase diagrams of textures and structures}

We start with an experimental phase diagram of cholesteric structures in the
homeotropic cells similar to the one reported in \cite%
{Oswald-review00,CHnegativeDeltaE} and then explore how this diagram is
affected by rubbing of homeotropic substrates, using different voltage
driving schemes, and introducing nucleation sites. We note that for pitch $%
p\gtrapprox 5\mu m$ and the cell gap $d\gtrapprox 5\mu m$ much larger than
the anchoring extrapolation length ($<1\mu m,$ describing polar anchoring at
the interface of CLC and JALS-204 layer), the observed structures depend on $%
\rho =d/p$ but not explicitly on $d$ and $p$. We therefore construct the
diagrams of structures in the plane of the applied voltage $U$ and\ the
frustration ratio $\rho $; to describe the phase diagram we adopt the
terminology introduced in Ref. \cite{Oswald-review00}. The diagrams display
director structures (phases) of homeotropic untwisted state, isolated $CF$s
and periodically arranged $CF$s, the $TIC$ and the modulated (undulating) $%
TIC$. The phase boundary lines are denoted as $V0-V3$, $V01$, $V02$, Fig.1,
similarly to Refs. \cite{Oswald-review00,CHnegativeDeltaE} (for comparison
with the phase diagrams reported for other LCs). As we show below, the phase
diagram can be modified to satisfy requirements for several electro-optic
applications of the CLC structures.

\subsubsection{Cells with unrubbed homeotropic substrates}

The diagram for unrubbed homeotropic cells is shown in Fig.1. The completely
unwound homeotropic texture is observed at small $U$ and $\rho $, Fig.1. At
high $U$ above $V0$, $V01$, and $V02$, the $TIC$ with some amount of
director twist (up to $2\pi $, helical axis along the cell normal) is
observed; the twist in $TIC$ is accompanied by splay and bend deformations.
The $TIC$ texture is homogeneous within the plane of a cell except that it
often contains the so-called umbilics, defects in direction of the tilt \cite%
{de Gennes-book}, Fig.2f. Periodically arranged $CF$s are observed for
voltages $U\approx 1.5-3.5Vrms$ and for $0.5<\rho <1$, Fig.1 and Fig.2b-e.
If the values of $U$ and $\rho $ are between the $V0$ and $V01$ boundary
lines,\ Fig.1, a transient $TIC$ appears first but then it is replaced
(within $0.1-10s$ after voltage pulse, depending on $\rho $ and $U$) by a
periodic pattern of $CF$s, which also undergoes slow relaxation; equilibrium
is reached only in $3-50s$, Fig.2d. The isolated $CF$s coexisting with the
homeotropic state are observed at $U\lessapprox 1.8$ and for $0.75<\rho <1$,
Fig.1 and Fig.2a,b. For $\rho $ and $U$ between $V0$ and $V1$, the isolated
fingers start growing from nucleation sites such as spacers, Fig.2b, or from
already existing fingers. In both cases the $CF$s separated by homeotropic
regions split in order to fill in the entire space with a periodic texture
of period $\sim p$, similar to the one shown in Fig.2c. In the region
between $V1$ and $V2$, isolated $CF$s nucleate and grow but they do not
split and do not fill in the whole sample; fingers in this part of diagram
coexist with homeotropic untwisted structure, Fig.2a. Hysteresis is observed
between $V2$ and $V3$ lines: a homeotropic texture is observed if the
voltage is increased, but isolated fingers coexisting with untwisted
homeotropic structure can be found if $U$ is decreased from the initial high
values. Even though the neighboring $CF$s in the fingers pattern are locally
parallel to each other, Fig.2c,d, there is no preferential orientation of
the fingers in the plane of the cell on the scales $\gtrapprox 10mm$.
Finally, the periodic structure observed between $V01$ and $V02$ does not
contain interspersed homeotropic regions, Fig.2e. The director field of $CF$%
s as well as other structures of the diagram will be revealed by FCPM below,
see Sec. III.B.

The behavior of the voltage-driven transitions between untwisted homeotropic
and different types of twisted structures is reminiscent of conventional
temperature-driven phase transitions with voltage playing a role similar to
temperature. The phase diagram of structures has a Landau tricritical point $%
\rho =\rho _{tricritical}$ at which $V2$ and $V0$ meet. The order of the
transition changes from the second order (continuous) at $\rho <\rho
_{tricritical}$ to the first order (discontinuous, proceeding via
nucleation) at $\rho >\rho _{tricritical}$, Fig.1. The phase diagram also
has a triple point at $\rho =\rho _{triple},$ where $V0$ and $V01$ meet. At
the triple point, the untwisted homeotropic texture coexists with two
different twisted structures, spatially uniform $TIC$ and periodic fingers
pattern. The phase diagram and transitions in homeotropic cells with
perpendicular easy axes at the substrates are qualitatively similar to those
reported by Oswald et al. \cite{Oswald-review00,CHnegativeDeltaE} for other
materials; both qualitative and quantitative differences are observed when
the homeotropic substrates are rubbed to produce slightly tilted easy axes
at the confining substrates as discussed below.

\subsubsection{Effects of rubbing and nucleation centers}

Rubbing of the homeotropic alignment layers induces small pretilt angle from
the vertical axis, $\gamma <2^{\circ }$. The azimuthal degeneracy is
therefore broken, and the projection of easy axis defines a unique direction
in the plane of a cell. Therefore, even rubbing of only one of the cell
substrates has a strong effect on the CLC structures: (1) no umbilics are
observed in the $TIC$, Fig.3a;\ (2) $CF$s preferentially align along the
rubbing direction, Fig.3b. In addition, the homeotropic-$TIC$ transition,
which is sharp in cells with vertical alignment, becomes somewhat blurred
for rubbed homeotropic substrates with small $\gamma $, Fig.3c,d.

In principle, one can set opposite rubbing directions on the substrates; we
report a phase diagram of structures for such anti-parallel rubbing in
Fig.4. The cells used to obtain the diagram were constructed from thick $3mm$
glass plates and only the mylar spacers at the cell edges were used to set
the cell gap thickness. Compared to phase diagrams of structures with
unrubbed substrates, dramatic changes are observed at $\rho \gtrapprox 0.5$.
The direct homeotropic to $TIC$ transition is observed up to $\rho \approx
0.75$. The experimental triple and tricritical points are closer to each
other than for unrubbed cells (compare \ Fig.1 and Fig.4). Interestingly,
within the range $0.6<\rho <0.75$ and upon increasing $U$, one first
observes a homogeneous $TIC$, Fig.5b, which is then replaced by a periodic
fingers pattern at higher $U$, Fig.5c, and again a uniform $TIC$ at even
higher $U>3-3.5V$, Fig.5d. The same sequence, $TIC$-fingers-$TIC$%
-homeotropic, is also observed upon decreasing $U$ from initial high values.
Pursuing the analogy with temperature-driven phase transitions, the $TIC$
texture between the fingers pattern and homeotropic texture can be
considered as a re-entrant $TIC$ phase. As compared to unrubbed cells, the
antiparallel rubbing has little effect on $V0$, but shifts the other
boundary lines towards increasing $\rho $. The effects of anti-parallel
rubbing on the phase diagram can be explained as follows. At $\rho \sim 0.5$
anti-parallel rubbing matches the director twist of $TIC,$ which at high $U$
is $\approx \pi $. Therefore, $TIC$ is stabilized by anti-parallel rubbing
and $CF$s do not appear until higher $\rho $, Fig.4.

The transient $TIC$ disappears if large quantities of spacers ($>100/mm^{2}$%
) or other nucleation sites for fingers are present in the cells with
anti-parallel rubbing; in this case the phase diagram is closer to the one
shown in Fig.1. The spacers with perpendicular surface anchoring produce
director distortions in their close vicinity even in the part of diagram
corresponding to homeotropic unwound state, Fig.6a. In the vicinity of the
homeotropic-$TIC$ transition, Fig.6b, the director realignment starts in the
vicinity of spacers. Similar to the observations in Refs. \cite%
{InvWalls,PECladis}, particles with perpendicular surface anchoring cause
inversion walls ($IW$s) and disclinations. The $TIC$ with $0.5<\rho <0.75$,
Fig.6c, is eventually replaced by fingers, which are facilitated by the
particles, Fig.6d. Moreover, even at high $U$, $TIC$ remains spatially
non-uniform and contains different types of $IW$s and disclination lines %
\cite{InvWalls,PECladis}, which are caused by the boundary conditions at the
surfaces of the particles.

\subsubsection{Phase diagrams for different voltage driving schemes}

The phase diagrams of structures shown in Figs.1 and 4 were obtained with
constant amplitude sinusoidal voltages applied to the cells.\ The diagram
changes dramatically if the applied voltage is modulated. The effect is
especially strong in the cells with rubbed homeotropic substrates, for which
we present results in Fig.7a-c; somewhat weaker effect is also observed for
unrubbed substrates. We explored modulation with rectangular-type,
triangular, and sinusoidal signals of different duration and modulation
depth. The strongest effect is observed with $100\%$ modulation depth and
sinusoidal modulation signal at frequencies (10-200)$Hz$. The fingers
patterns are shifted towards increasing $\rho $, Fig.7. At the same time,
the $rms$ voltage values of homeotropic-$TIC$ transition are practically the
same for different voltage driving schemes, Figs.1,4,7. We assume that the
effect of amplitude modulation is related to the very slow dynamics of some
of the structures (see Secs. III.A.1, III.A.2), such as $CF$s; the
corresponding parts of the diagram are the most sensitive to voltage driving
schemes.

The substantial combined effect of rubbing and voltage driving schemes is
important for practical applications of the homeotropic-$TIC$ transition
when it is important to have strongly twisted but finger-free field-on state %
\cite{EyeLC1,Bistable-chiral}. We therefore present only the diagrams
corresponding to the largest $\rho $ values at which fingers do not appear
for given surface rubbing conditions, Fig.7. On the other hand, voltage
modulation could be a way to study the stability of different parts of the
diagram in the $\rho ,U$ plane and deserves to be explored in more details;
we leave this for forthcoming publications. Finally, to understand the
diagrams and transitions explored in this section it is important to know
the director fields that are behind different textures; this will be
explored in the following section.

\subsection{Director structures}

\subsubsection{Spatially-homogeneous twisted structures, umbilics, and
inversion walls}

In this section, we take advantage of the FCPM and study the director field $%
{\bf \hat{n}}\left( x,y,z\right) $ in the vertical cross-sections (i.e.,
along the $z$, normal to the cell substrates) of the cholesteric structures.
This is important as, for example, in the $TIC$\ ${\bf \hat{n}}$ varies only
along $z$ and not in the plane of a cell. The $TIC$, observed above the $V0$
and $V02$ lines in the phase diagrams, Figs.1,4,7, can be visualized as
having ${\bf \hat{n}}$ rotating with distance from the cell wall on a cone
whose axis is along $z$; the half angle of this cone varies from $\theta =0$
at the substrates to $\theta _{\max }$ in the middle plane of a cell ($%
\theta _{\max }<\pi /2$), Fig.8. FCPM reveals that the in-plane twist of the
director in the $TIC$ depends on $\rho $. For small $\rho \approx 0,$ the $%
TIC$ contains practically no in-plane twist. When $\rho \approx 1/2$, the
in-plane twist at high $U$ reaches $\pi $, Fig.8a,c. Finally, when $\rho
\approx 1$, the twist of the $TIC$ structure at high $U$ can reach $2\pi $,
Fig.8b,d. The maximum in-plane twist at high $U$ is $\approx 2\pi \rho $; we
stress that the twist of $TIC$ depends not only on $\rho $, but also on $U.$
In addition, for the cells with rubbed homeotropic plates, the in-plane
twist is affected by the rubbing direction. For example, the re-entrant $TIC$
in the rubbed cells of $0.5<\rho <0.75$ has twist $\approx \pi $ at small $U$
just above the $V0$ and the twist close to $2\pi $ at high voltages $U>4Vrms$%
. If both of the homeotropic substrates are rubbed, the natural twist of the 
$TIC$ structure may or may not be compatible with the tilted easy axes at
the confined substrates. Since the amount of twist in the $TIC$ depends both
on $\rho $ and $U$, it is impossible to match tilted homeotropic boundary
conditions to a broad range of $\rho $ and $U$. However, since the in-plane
anchoring is weak, the effect of rubbing on the twist in $TIC$ is not as
strong as in the case of planar cells.

The FCPM also allows us to probe the defects that appear in $TIC$. We
confirm that the defects with four brushes, Fig.2f, are umbilics of strength 
$\pm 1$\cite{de Gennes-book} rather than disclinations with singular cores.
We also verify that the umbilics are caused by degeneracy of director tilt
when $U$ is applied; such degeneracy is eliminated by rubbing, Fig.3a.
Within $TIC$, we also observe $IW$s\cite{InvWalls,PECladis}. The appearance
of these walls was previously attributed to a variety of factors, such as
flow of liquid crystal, hydrodynamic instabilities, alignment induced at the
edges of the sample, and others \cite{InvWalls,PECladis}. FCPM\ observations
indicate that in the presence of spacers with perpendicular anchoring, the $%
IW$s appear at the particles when $U$ above the threshold for homeotropic-$%
TIC$ transition is applied. This is believed to be caused by director
distortions in the vicinity of the particles \cite{InvWalls,PECladis}. When
the confinement ratio is $0.5<\rho <0.75$, the distorted $TIC$ with
umbilics, disclinations, and $IW$s is replaced by $CF$s with the spacers
serving as nucleation sites for the fingers, Fig.6.

\subsubsection{Fingers structures; nonsingular fingers of $CF1$-type}

Fingers structures have translational invariance along their axes ($y$%
-direction) and can be observed as isolated between $V3$ and $V0$ or
periodically arranged between $V0$ and $V01$ boundary lines, Figs.1,4,7 (see
Sec.III.A for details). We again take advantage of FCPM by visualizing the
vertical cross-sections and then reconstruct ${\bf \hat{n}}\left(
x,y,z\right) $ of the fingers directly from the experimental data. To
describe the results, we use the $CFs$ classification of Oswald et al. \cite%
{Oswald-review00}. The finger of $CF1$ type is the most frequently observed
in cells with vertical as well as slightly tilted alignment, Fig.9. $CF1$ is
isolated and co-existing with the homeotropic untwisted structure between $%
V3 $ and $V0$ and is a part of the spatially periodic pattern between $V0$
and $V01$ lines, Figs.1,4,7. The reason for abundance of $CF1$, is that it
can form from $TIC$ by continuous transformation of director field above the 
$V0$ line and it also can easily nucleate from homeotropic untwisted
structure below the $V0$ boundary line, Figs.1,4,7\cite{Oswald-review00}.

The director structure of $CF1$ reconstructed from the FCPM vertical
cross-section, Fig.9, is in a good agreement with the results of computer
simulations \cite%
{Oswald-review00,LGil,ourreview,PressArrott,MCLC-PressArrott}. In the $CF1$,
the axis of cholesteric twist is tilted away from the cell normal $z$,
Fig.9; the in-plane twist in direction perpendicular to the finger in the
middle of a cell is $2\pi $. The isolated $CF1s$ that are separated from
each other by large regions of homeotropic texture, Fig.2a, assume random
tilt directions. The width of an isolated $CF1$ is somewhat larger than $d$;
this is in a good agreement with computer simulations of L. Gil \cite{LGil}. 
$CF1$ is nonsingular in ${\bf \hat{n}}$ (i.e., the spatial changes of ${\bf 
\hat{n}}$ are continuous and ${\bf \hat{n}}$ can be defined everywhere
within the structure) as the twist is accompanied with escape of ${\bf \hat{n%
}}$ into the third dimension along its center line. An isolated $CF1$ can be
represented as a quadrupole of the non-singular $\lambda $-disclinations,
two of strength $+1/2$ and two of strength $-1/2$, as shown in Fig.9b. The $%
\lambda $-disclinations, with core size of the order of $p$, cost much less
energy than the disclinations with singular cores \cite%
{KlemanLavrentovichBook}. The pair of disclinations $\lambda ^{+1/2}\lambda
^{-1/2}$ introduces $2\pi $-twist at one homeotropic substrate; this $2\pi $%
-twist is then terminated by introducing another $\lambda ^{+1/2}\lambda
^{-1/2}$ pair in order to satisfy the homeotropic boundary conditions at
another substrate, Fig.9b. A segment of an isolated $CF1$ has different
ends; one is rounded while the another is pointed.\ Behavior of these ends
is different during growth; the pointed end remains stable, while the
rounded end continuously splits, as also discussed in Ref. \cite{Oswald2004}.

The FCPM vertical cross-section, Fig.10,\ reveals details of $CF1$ tiling
into periodically arranged structures that are observed above $V0$ line,
Figs.1,4,7. When $U$ or $\rho $ are relatively large, the $CF1$ fingers are
close to each other so that the homeotropic regions in between cannot be
clearly distinguished. The tilt of the helical axis in the periodic $CF$
structures is usually in the same direction, Fig.10. A possible explanation
is that the elastic free energy is minimized since the structure of
unidirectionally tilted $CFs$ is essentially space-filling.\ On rare
occasions, the tilt direction of neighboring $CF1s$ is opposite. Upon
increasing $U$, the width of fingers originally separated by homeotropic
regions, Fig.11a, gradually increases, Fig.11b-e; the fingers then merge to
form a periodically modulated $TIC$, Fig.11f. Finally, at high applied
voltages, the transition to uniform $TIC$ is observed, Fig.11g. The details
of transformation of periodically arranged fingers into the in-plane
homogeneous $TIC$ via the modulated (undulating) twisted structure were not
known before and would be difficult to grasp without FCPM. $TIC$ can also be
formed by expanding one of the $CF1$s; structure of coexisting fingers and $%
TIC$ contains only $\lambda $-disclinations nonsingular in ${\bf \hat{n}}$
again demonstrating the natural tendency to avoid singularities, Fig.12.
Periodically arranged $CF1$s slowly (depending on rubbing, $U$ and $\rho $;
usually up to $1s$) appear from $TIC$ if $U$\ is between $V0$ and $V01,$ and
quickly disappear (in less than $50ms$) if $U$ is increased above $V02$.
This allowed us to use the amplitude-modulated voltage driving schemes in
combination with rubbing and obtain finger-free $TIC$ up to $\rho \approx
0.8 $ (Sec. III.A.3), as needed for applications of $TIC$ in the
electrically driven light shutters, intensity modulators, eyewear with
tunable transparency, displays, etc. \cite%
{Blinov-Ch-book,EyeLC1,Patel,Bistable-chiral}

\subsubsection{Fingers of $CF2$, $CF3$, and $CF4$ types containing defects;
T-junctions of fingers}

Another type of fingers is $CF2$, Fig.13, which is observed for vertical as
well as slightly tilted alignment in the same parts of the diagram as $CF1$,
Figs.1,4,7. However, in contrast to the case of nonsingular $CF1$, a segment
of $CF2$ has point defects at its ends. Because of this, $CF2$ does not
nucleate from the homeotropic or $TIC$ structures as easily as $CF1$ and
normally dust particles, spacers, or irregularities are responsible for its
appearance. Therefore, fingers of $CF2$-type, Fig.13, are found less
frequently than $CF1$. Using FCPM, we reconstruct the director structure in
the vertical cross-section of $CF2$, Fig.13; the experimental result
resembles the one obtained in computer simulations by Gil and Gilli \cite%
{Gil-PRL98}, proving the latest model of $CF2$ \cite%
{Oswald-review00,Gil-PRL98} and disproving the earlier ones \cite%
{Baudry-PRE99}. Within the $CF2$ structure, one can distinguish the
non-singular $\lambda ^{+1}$ disclination in the central part of the cell
and two half-integer $\lambda ^{-1/2}$ disclinations in the vicinity of the
opposite homeotropic substrates. The total topological charge of the $CF2$
is conserved, similarly to the case of $CF1$.\ Polarizing microscopic
observations show that unlike in $CF1$-type fingers, the ends of $CF2$
segments have similar appearance. FCPM reveals that the point defects (of
strength $1$) at the two ends have different locations being closer to the
opposite substrates of a cell. Unlike the $CF1$ structure, $CF2$ is not
invariant by $\pi $-rotation around the $y$-axis along the finger, as also
can be seen from the FCPM cross-section, Fig.13. Different symmetries of $%
CF1 $ and $CF2$ are responsible for their different dynamics under electric
field \cite{Oswald-review00,Gil-PRL98}. This, along with computer
simulations, allowed Gil and Gilli \cite{Gil-PRL98} to propose the model of $%
CF2$; our direct imaging using FCPM unambiguously proves that this model is
correct, Fig.13b.

The isolated $CF2$ fingers (coexisting with homeotropic state)\ expand, when 
$U\gtrapprox 2.1Vrms$ is applied, Fig.13c,d. The structures with
non-singular $\lambda $-disclinations often separate the parts of a cell
with different twist, Fig.14; they resemble the structures of thick lines
that are observed in Grandjean-Cano wedges with planar surface anchoring %
\cite{SmalyukPRE,SmalyukhPRL}. The appearance of these lines in homeotropic
cells is facilitated by sample thickness variations and spacers. The width
of $CF2$ coexisting with the homeotropic state is usually the same or
somewhat (up to 30\%) smaller than $CF1$; this can be seen in Fig.15 showing
a $T$-junction of the $CF1$ and $CF2$ fingers. Even though $CF1$ and $CF2$
have similar appearance under a polarizing microscope \cite{Oswald-review00}%
, FCPM allows one to clearly distinguish these structures.\ Note also the
tendency to avoid singularities in ${\bf \hat{n}}$ evidenced by the
reconstructed structure of the $T$-junction, Fig.15b.

The metastable cholesteric fingers of $CF3$-type, Fig.16, occur even less
frequently than $CF2s$. The director structure of this finger was originally
proposed by Cladis and Kleman \cite{CladisKleman}. In polarizing microscopy
observations, the width of $CF3$ fingers is about half of that in $CF1$ and $%
CF2$. The reconstructed FCPM\ structure of $CF3$ indicates that the director 
${\bf \hat{n}}$ rotates through only $\pi $ along the axis perpendicular to
the finger ($x$-axis). This differs from both $CF1$ and $CF2$, which both
show a rotation of ${\bf \hat{n}}$ through $2\pi $. \ Two twist
disclinations of opposite signs near the substrates allow the cholesteric $%
\pi $-twist in the bulk to match the homeotropic boundary conditions,
Fig.16. The structure of $CF3$ is singular in ${\bf \hat{n}}$; the
disclinations are energetically costly and this explains why $CF3$ is
observed rarely even in the cells with vertical alignment. In cells with
rubbed homeotropic substrates $CF3$ was never observed. This is likely due
to the easy axis having the same tilt on both sides of the finger on a
rubbed substrate,\ whereas the $\pi $-twisted configuration of \ $CF3$
requires director tilt in opposite directions.

The $CF4$-type metastable finger shown in Fig.17 is also singular, and is
usually somewhat wider than the other $CFs$. It can be found in all regions
of existence of $CF1$, Fig.1, but is very rare and usually is formed after
cooling the sample from isotropic phase. $CF4$ contains two singular
disclinations at the same substrate. In the plane of a cell, the director $%
{\bf \hat{n}}$ rotates by $2\pi $ with the twist axis being along ${\bf x}$
and perpendicular to the finger. Using the direct FCPM imaging, we
reconstruct the director structure of $CF4$, Fig.17b, which is in a good
agreement with the model of Baudry et al. \cite{T-Junktions}. The bottom
part of this finger, Fig.17, is nonsingular, and is similar in this respect
to $CF1$ and $CF2$; its top part, however, contains two singular twist
disclinations. The $CF4$ structure is observed only in cells with no
rubbing. Similar to the case of $CF3$, the structure of $CF4$ is not
compatible with uniform tilt produced by rubbing of homeotropic substrates.
Of the four different fingers structures, $CF4$ might be the least favorable
energetically, since it usually rapidly transforms into $CF1$ or $CF2$; less
frequently, it also splits into two $CF3$ fingers. Transformation of other
fingers into $CF4$ was never observed.

\section{Discussion}

The system that we study is fairly rich and complicated; some of the
structures and transitions can be described analytically while the other
require numerical modeling. Below we first restrict ourselves to
translationally uniform structures (i.e., homeotropic and $TIC$) which can
be described analytically. We then discuss the other experimentally observed
structures and transitions comparing them to the analytical as well as
numerical results available in literature \cite%
{Oswald-review00,CHnegativeDeltaE}, as well as our own numerical study of
the phase diagram that will be published elsewhere\cite{Garlandetal}.
Finally, we discuss the practical importance of the obtained results on the
phase diagrams of director structures.

\subsection{Translationally uniform homeotropic and $TIC$ structures}

We represent the director ${\bf \hat{n}}$ in terms of the polar angle $%
\theta $ (between the director and the $z$-axis) and the azimuthal angle $%
\phi $ (the twist angle); $\psi $ is electric potential. For the $TIC$
configurations, these fields are functions of $z$ only, and the Oseen-Frank
free-energy density takes the form 
\begin{eqnarray}
2f &=&(K_{11}\sin ^{2}\!\theta +K_{33}\cos ^{2}\!\theta )\theta
_{z}^{2}+(K_{22}\sin ^{2}\!\theta +K_{33}\cos ^{2}\!\theta )\sin
^{2}\!\theta \,\phi _{z}^{2}  \nonumber \\
&&{}-K_{22}\frac{4\pi }{p}\sin ^{2}\!\theta \,\phi _{z}+K_{22}\frac{4\pi ^{2}%
}{p^{2}}-(\varepsilon _{\perp }\sin ^{2}\!\theta +\varepsilon _{\parallel
}\cos ^{2}\!\theta )\psi _{z}^{2},  \label{eqn:fe}
\end{eqnarray}%
where, $K_{11}$, $K_{22}$, and $K_{33}$ are the splay, twist, and bend
elastic constants, respectively; $\varepsilon _{\parallel }$,$\varepsilon
_{\perp }$ are the dielectric constants parallel and perpendicular to ${\bf 
\hat{n},}$ respectively; $\theta _{z}=d\theta /dz,$ $\phi _{z}=d\phi /dz,$
and $\psi _{z}=d\psi /dz$. The associated coupled Euler-Lagrange equations
are 
\begin{eqnarray}
\lefteqn{\frac{d}{dz}\left[ \left( K_{11}\sin ^{2}\!\theta +K_{33}\cos
^{2}\!\theta \right) \theta _{z}\right] =\sin \theta \cos \theta
\Bigl\{(K_{11}-K_{33})\theta _{z}^{2}}  \nonumber \\
&&{}+\left[ (2K_{22}-K_{33})\sin ^{2}\!\theta +K_{33}\cos ^{2}\!\theta %
\right] \phi _{z}^{2}-K_{22}\frac{4\pi }{p}\phi _{z}-\Delta \varepsilon \psi
_{z}^{2}\Bigr\},  \label{eqn:ELa}
\end{eqnarray}

\begin{equation}
\frac{d}{dz}\left\{ \sin ^{2}\!\theta \left[ (K_{22}\sin ^{2}\!\theta
+K_{33}\cos ^{2}\!\theta )\phi _{z}-K_{22}\frac{2\pi }{p}\right] \right\} =0,
\label{eqn:ELb}
\end{equation}%
\begin{equation}
\frac{d}{dz}\left[ \left( \varepsilon _{\perp }\sin ^{2}\!\theta
+\varepsilon _{\parallel }\cos ^{2}\!\theta \right) \psi _{z}\right] =0,
\label{eqn:ELc}
\end{equation}%
with associated boundary conditions $\theta (0)=\theta (d)=0$; $\phi
(0),\phi (d)$ undefined; and $\psi (0)=0$, $\psi (d)=U$. Dielectric
anisotropy is negative for the studied material, $\Delta \varepsilon
=\varepsilon _{\parallel }-\varepsilon _{\perp }<0$. Representative
solutions of these equations are plotted in Fig.18 and describe how $\theta
(z),$ $\phi (z),$ and $\psi (z)$ vary across the cell.

Equations (\ref{eqn:ELb}) and (\ref{eqn:ELc}) above admit first integrals, 
\begin{equation}
\phi _{z}=\frac{K_{22}}{K_{22}\sin ^{2}\!\theta +K_{33}\cos ^{2}\!\theta }%
\frac{2\pi }{p}  \label{eqn:phiz}
\end{equation}%
and 
\begin{equation}
\left( \varepsilon _{\perp }\sin ^{2}\!\theta +\varepsilon _{\parallel }\cos
^{2}\!\theta \right) \psi _{z}=\frac{U}{\displaystyle\int_{0}^{d}\frac{1}{%
\varepsilon _{\perp }\sin ^{2}\!\theta +\varepsilon _{\parallel }\cos
^{2}\!\theta }dz},
\end{equation}%
which allow to express the free energy in terms of the tilt angle $\theta $
only: 
\begin{eqnarray}
{\cal F}[\theta ] &=&\frac{1}{2}\int_{0}^{d}\left[ (K_{11}\sin ^{2}\!\theta
+K_{33}\cos ^{2}\!\theta )\theta _{z}^{2}+\frac{K_{2}K_{3}\cos ^{2}\!\theta 
}{K_{22}\sin ^{2}\!\theta +K_{33}\cos ^{2}\!\theta }\frac{4\pi ^{2}}{p^{2}}%
\right] dz  \nonumber \\
&&{}-\frac{1}{2}U^{2}\left[ \int_{0}^{d}\frac{1}{\varepsilon _{\perp }\sin
^{2}\!\theta +\varepsilon _{\parallel }\cos ^{2}\!\theta }dz\right] ^{-1}.
\end{eqnarray}%
This is similar to Eq. (3.221) of Ref. \cite{stewart:04} (see p. 91), where
the splay Freedericksz transition with a coupled electric field is
discussed. We expand the free energy in terms of $\theta (z)$ about the
undistorted $\theta =0\,$homeotropic configuration to obtain%
\begin{equation}
{\cal F}[\theta ]=\frac{1}{2}\left[ \frac{4\pi ^{2}dK_{22}}{p^{2}}-\frac{%
\varepsilon _{\parallel }U^{2}}{d}\right] +\frac{1}{2}\int_{0}^{d}\left[
K_{33}\theta _{z}^{2}-\left( \frac{\Delta \varepsilon U^{2}}{d^{2}}+\frac{%
4\pi ^{2}K_{22}^{2}}{p^{2}K_{33}}\right) \theta ^{2}\right] dz+O(\theta
^{4}).
\end{equation}%
The first term is the free energy of the uniform homeoptropic configuration.
The second-order term is positive definite if $U$ and $\rho $ are
sufficiently small. Ignoring higher order terms, we find that the loss of
stability occurs when 
\begin{equation}
\frac{4K_{22}^{2}}{K_{33}^{2}}\rho ^{2}+\frac{\Delta \varepsilon }{K_{33}\pi
^{2}}U^{2}=1.  \label{eqn:ellipse}
\end{equation}%
Eq. (\ref{eqn:ellipse}) is the spinodal ellipse. The homeotropic
configuration is metastable with respect to translationally homogeneous
perturbations for the $\rho $ and $U$\ parameters inside the ellipse (\ref%
{eqn:ellipse}), which corresponds to the boundary line $V0$, Figs.1,4,7. Eq.
(\ref{eqn:ellipse}) gives the threshold voltage for transition between
homeotropic and $TIC$ structures:

\begin{equation}
U_{th}=\pi \sqrt{K_{33}/\Delta \varepsilon }\cdot \sqrt{1-4\rho
^{2}K_{22}^{2}/K_{33}^{2}}.  \label{U0-threshold}
\end{equation}

Eq. (\ref{U0-threshold}) is in a good agreement with our experimental
results described in Sec. III above and with Ref. \cite{Rosenblatt}. The
experimental data for boundary line $V0$ are well described by Eq. (\ref%
{eqn:ellipse}) for rubbed and unrubbed homeotropic cells, Figs.1,4,7.
According to the linear stability analysis above, the ellipse in the $\rho $-%
$U$ plane (\ref{U0-threshold}) defines the limit of metastability of the
homeotropic phase: for $\rho $ and $U$ inside this ellipse, the uniform
homeotropic configuration is metastable, while outside the ellipse, it is a
locally unstable equilibrium. In an idealized cell with infinitely strong
homeotropic anchoring and no pretilt, the transition from homeotropic to $%
TIC $ is a forward pitchfork bifurcation, that is, a second-order
transition. For voltages $U$ below the $V0$ line (inside the ellipse), the $%
TIC$ configuration does not exist. On the other hand, when there is a slight
tilt of the easy axis, the reflection symmetry is broken. The pitchfork is
unfolded into a smoother transition, i.e., the transition becomes
supercritical and the precise transition threshold is not well defined. The
experimentally observable artefact of this is the somewhat blurred
transition, which is described in Sec.~III.A.2, for the cells with rubbing
and resembles a similar effect in planar cells with small pretilt\cite%
{Yehbook}.

The above analysis allows one to understand the dependence of the total
in-plane twist of $TIC$ on $\rho $ and $U$ that was described in
Sec.~III.B.1. The first integral (\ref{eqn:phiz}) gives the tilt-dependence
of the local twist rate. The total twist across a cell of thickness $d$ is 
\begin{equation}
\Delta \phi =\frac{2\pi }{p}\int_{0}^{d}\frac{K_{22}}{K_{22}\sin
^{2}\!\theta +K_{33}\cos ^{2}\!\theta }\,dz.  \label{eqn:Dphi}
\end{equation}%
For the AMLC-0010 material with $K_{22}/K_{33}\approx 0.42$ for given $\rho $
the total twist $\Delta \phi $ can be varied 
\begin{equation}
0.42\ast 2\pi \rho <\Delta \phi <2\pi \rho ,  \label{eqn:DphiLimits}
\end{equation}%
by changing $U$. $\Delta \phi $\ approaches the lower limit for relatively
small $U$\ that are just above $U_{th}$ and $\theta \approx 0$ and the upper
limit for $U\gg U_{th}$ and $\theta \approx \pi /2$ . This analysis is in a
good agreement with the FCPM images of the vertical cross-sections of $TIC$
for different $\rho $, as described in Sec. III. Finally, knowledge of $%
\Delta \phi $ variation with changing $U$ is important for the practical
applications of $TIC$ as it will be discussed below.

\subsection{Other structures and transitions of the phase diagram}

Modeling of transitions associated with $CFs,$ in which ${\bf \hat{n}}$ is a
function of two coordinates, is more complicated than in the case of $TIC$.
Ribi\`{e}re, Pirkl, and Oswald\cite{CHnegativeDeltaE} obtained complete
phase diagram in calculations assuming a simplified model of a cholesteric
finger. This theoretical diagram qualitatively resembles our experimental
result for the cells with vertical alignment, Fig.1. We explored the phase
diagram using 2-D numerical modeling in which the equilibrium structure of
the $CFs$ and equilibrium period of periodically arranged fingers are
determined from energy minimization in a self-consistent way and the
nonlocal field effects are taken into account\cite{Garlandetal}. The
numerical phase diagrams show a good quantitative agreement with the
experimental results presented here, predicting even the re-entrant behavior
of $TIC$ that we experimentally obtain for cells with rubbed substrates
(Sec.III.A.2). Presentation of these results requires detailed description
of numerical modeling and will be published elsewhere\cite{Garlandetal}.
Therefore, we only briefly discuss the qualitative features of the phase
diagrams shown in Figs.1,4,7 in the light of the previous theoretical studies%
\cite{Oswald-review00,CHnegativeDeltaE} and also summarize the new
experimental results below.

The important feature of the studied diagram is that the nematic-cholesteric
transition changes order: it is second order for $0<\rho <\rho
_{tricritical} $ and first order for $\rho >\rho _{tricritical}$, Fig.1, in
agreement with Refs.\cite{Oswald-review00,CHnegativeDeltaE} The phase
diagram has a triple point at $\rho =\rho _{triple}$, where $V0$ and $V01$
meet and the untwisted homeotropic texture coexists with two twisted
structures, $TIC$ and periodically arranged $CFs$, Figs1,4,7. For vertical
alignment, the direct voltage-driven homeotropic-$TIC$ transition is
observed at small $\rho \lessapprox 0.5$. Structures of isolated $CFs$ and
periodically arranged $CFs $ occur for $0.5\lessapprox \rho <1$ and
intermediate $U$ between the homeotropic state and $TIC$. The theoretical
analysis of Ref. \cite{CHnegativeDeltaE} allows one to determine $\rho $
corresponding to the tricritical and triple points in the phase diagrams.
Solving the equations given in Ref.\cite{CHnegativeDeltaE} numerically\cite%
{error} \ and using the material parameters of the AMLC-0010 host doped with
the chiral agent ZLI-811, we find $\rho $ corresponding to the triple and
tricritical points: $\rho _{triple}=0.816$ and $\rho _{tricritical}=0.861.$
These values are somewhat larger than $\rho _{triple}$ and $\rho
_{tricritical}$ determined experimentally for the cells with vertical
alignment, Figs.1,4,7, as also observed in \cite{CHnegativeDeltaE} for other
CLCs. The calculated $\rho _{triple}$ and $\rho _{tricritical}$ are closer
to the experimental ones in the case of rubbed substrates; this may indicate
the possible role of umbilics and $IWs$ in the $TIC,$ which were not taken
into account in the model \cite{CHnegativeDeltaE} (umbilics and $IWs$ are
nucleation sites for fingers and may also increase elastic energy of $TIC$).
Agreement is improved when phase diagrams are obtained using 2-D numerical
modeling \cite{Garlandetal}. An interesting new finding revealed by the FCPM
is that upon increasing $U$ the periodically arranged fingers merge with
each other forming modulated (undulating) $TIC$ that is observed in a narrow
voltage range between the structures of $TIC$ and periodically arranged $CFs$%
, Figs.1,4,7. We also find that the phase boundaries can be shifted in a
controlled way by rubbing-induced tilt ($<2^{\circ }$) of easy axis from the
vertical direction, by introducing particles that become nucleation sites
for $CFs,$ as well as by using different amplitude-modulated voltage schemes.

A novel and unexpected result is the re-entrant behavior of $TIC$ in the
rubbed cells with $0.6\lessapprox \rho \lessapprox 0.75,$ which, however,
disappears if nucleation sites are present. FCPM allows us to directly and
unambiguously determine the 3-D director structures corresponding to
different parts of the phase diagram. In particular, we unambiguously
reconstruct the structures of four types of $CFs$. In all parts of diagrams
corresponding to stability or metastability of $CFs$, the fingers of $CF1$%
-type are the most frequently observed. $CF2$ fingers are less frequent; the
metastable $CF3$ and $CF4$ are very rare. Such findings indicate that
fingers of $CF1$-type have the lowest free energy out of four fingers; this
is consistent with the reconstruction of the structure of $CF1,$ which is
nonsingular in ${\bf \hat{n}}$. It is also natural that $CF2$ with singular
point defects and especially metastable $CF3$ and $CF4$ with singular line
defects are less frequently observed. In the case of rubbed homeotropic
substrates, only $CF1$ and $CF2$ are observed whereas $CF3$ and $CF4$ newer
appear because the rubbing-induced tilting of the easy axis at one or both
substrates contradicts with their symmetry.

\subsection{Control of phase diagrams to enable practical applications}

The combination of rubbing and amplitude-modulated voltage driving allows
one to suppress appearance of fingers up to high $\rho \approx 0.75$,
compare Fig.1 and Fig.7. This is a valuable finding for many practical
applications such as eyeglasses with voltage-tunable transparency and light
shutters,\cite{EyeLC1} bistable\cite{Bistable-chiral} and inverse twisted
nematic displays,\cite{Patel} etc. In these applications of the homeotropic-$%
TIC$ transition, it is important to have a broad range of well controlled
total twist $\Delta \phi $ in the finger-free $TIC$. The broad range of
voltage-tuned $\Delta \phi $ allows one to control optical phase retardation
in the displays and electro-optic devices\cite{Patel,Bistable-chiral} as
well as light absorption when the dye-doped CLC is used in the tunable
eyeglasses and light shutters\cite{EyeLC1}. A very subtle tilt of easy axis
from the vertical direction not only makes the director twist in $TIC$ vary
in a controlled way but also suppresses the appearance of fingers, $IW$s,
and umbilics, Figs.3-7. Slow appearance of fingers from $TIC$ and untwisted
homeotropic states allowed us to magnify the effect of rubbing via using
amplitude-modulated voltage schemes and suppress appearance of fingers up to
even higher $\rho $, Fig.7. For example, we can control $\Delta \phi
=55^{\circ }-270^{\circ }$ in the finger-free $TIC$. Even stronger effects
of rubbing and voltage modulation can be expected if the tilt of the easy
axis is increased. This might be implemented by using the approach recently
developed by Huang and Rosenblatt \cite{Homeotrop-tilt} in which case a tilt
of easy axis up to $30^{\circ }$ could be achieved. On the other hand, when
constructing cells for all of the above applications of tightly-twisted $TIC$%
, it is important to remember about the effect of particles, which become
nucleation sites for fingers and can cause their appearance at lower $\rho $%
. Such particles are often used as spacers to set cell thickness and it is,
therefore, important to either avoid their usage or limit (optimize) their
concentration in order to obtain finger-free $TIC$.

The finding that fingers align along the rubbing direction, Sec. III.A, may
enable the use of periodically-arranged $CF$s in switchable diffraction
gratings with the diffraction pattern corresponding to the field-on state.
The spatial periodicity and the diffraction properties of such gratings can
be easily controlled by selecting proper pitch $p$ and cell gap $d$, which
can be varied from sub-micron to tens of microns. Our preliminary study
shows that the grating periodicity can be changed in range $1-50\mu m$. More
detailed studies of the cholesteric diffraction gratings based on
voltage-induced well-oriented pattern of $CF$s will be published elsewhere.

\section{Conclusions}

The major findings of this work are threefold: (1) we obtained phase diagram
of CLC structures as a function of confinement ratio $\rho =d/p$ and voltage 
$U$ for different extra parameters such as rubbing, voltage driving,
presence of nucleation sites; (2) we enabled new applications of finger-free
tightly twisted $TIC$ and well-oriented fingers; (3) we unambiguously
deciphered 3-D director fields associated with different structures and
transitions in CLCs using FCPM. In the phase diagram, the direct homeotropic-%
$TIC$ transition upon increasing $U$ was observed for $\rho \lessapprox 0.5$%
; the analytical model of this transition is in a very good agreement with
the experiment. Structures of isolated and periodically arranged fingers
were found at $0.5\lessapprox \rho <1$ and intermediate $U$ between the
homeotropic and $TIC$ phases. We observed the re-entrant behavior of $TIC$
in the rubbed cells of $0.6\lessapprox \rho \lessapprox 0.75$ for which the
following sequence of transitions has been observed upon increasing $U$: (1)
homeotropic untwisted - (2) $TIC$ - (3) periodically arranged fingers - (4) $%
TIC$ with larger in-plane twist. The re-entrant behavior of $TIC$ is also
observed in our numerical study of the phase diagram that will be published
elsewhere\cite{Garlandetal}. The re-entrant $TIC$ disappears if nucleation
sites are present or amplitude-modulated driving schemes are used. Rubbing
also eliminates non-uniform in-plane structures such as umbilics and
inversion walls that otherwise are often observed in $TIC$ and also
influence the phase boundary lines. The lowest $\rho $ for which
periodically arranged fingers start to be observed upon increasing $U$ can
be shifted for up to $0.3$ towards higher $\rho $-values by rubbing and/or
voltage driving schemes. The FCPM allowed us to unambiguously determine and
confirm the latest director models \cite{Oswald-review00} for the vertical
cross-sections of four types of $CFs$ ($CF1-CF4$) while disproving some of
the earlier models\cite{Baudry-PRE99}. The $CF1$-type fingers are observed
in all regions of the phase diagrams where the fingers are either stable or
metastable; other fingers appear in the same parts of the diagram but less
frequently. For the rubbed cells, only two types of $CFs$ ($CF1$ and $CF2$)
are observed, which align along the rubbing direction. The new means to
control structures in CLCs are of importance for potential applications,
such as switchable gratings based on periodically arranged $CF$s and eyewear
with tunable transparency based on $TIC$.

\section{Acknowledgements}

The work is part of the AlphaMicron/TAF collaborative project ``Liquid
Crystal Eyewear'',\ supported by the State of Ohio and AlphaMicron, Inc.
I.I.S. and O.D.L. acknowledge support of the NSF, Grant DMR-0315523. I.I.S.
acknowledges Fellowship of the Institute for Complex and Adaptive Matter.
E.C.G. acknowledges support under NSF Grant DMS-0107761, as well as the
hospitality and support of the Department of Mathematics at the University
of Pavia (Italy) and the Institute for Mathematics and its Applications at
the University of Minnesota, where part of work was carried out. We are
grateful to M. Kleman, L. Longa, Yu. Nastishin, and S. Shiyanovskii for
discussions.

Figure Captions:

FIG.1. Phase diagram of structures in the $U-\rho $ parameter space for CLCs
in cells with homeotropic surface anchoring. The boundary lines $V0-V3$, $%
V01 $, $V02$ separate different phases (cholesteric structures). The two
dashed vertical lines mark $\rho _{triple}=0.816$ and $\rho
_{tricritical}=0.861$ as estimated according to Ref. \cite{CHnegativeDeltaE}
for the material parameters of AMLC-0010 - ZLI-811 LC mixture. The solid
line $V0$ was calculated using Eq.(1) and parameters of the used CLC; the
solid lines $V1-V3$, $V01$, $V02$ connect the experimental points to guide
the eye.

FIG.2. Polarizing microscope textures observed in different regions of the
phase diagram of structures shown in Fig.1: (a) isolated $CFs$ coexisting
with the homeotropic untwisted state between the boundary lines $V1$ and $V2$
of Fig.1; (b) dendritic-like growth of $CFs$ (observed between the boundary
lines $V0$ and $V1$); (c) branching of $CFs$ with increasing voltage,
between the boundary lines $V0$ and $V1$; (d) periodically arranged $CFs$
where the individual $CFs$ are separated by homeotropic narrow stripes,
observed between $V0$ and $V01$; (e) $CFs$ merge producing undulating $TIC$,
observed between $V01$ and $V02$; (f) $TIC$ with umbilics, observed above
the lines $V0$ and $V02$. Picture shown in part (b) was taken about 2
seconds after voltage was applied; it shows an intermediate state in which
the circular domains grow from nucleation sites and will eventually fill in
the whole area of the cell by the fingers.

FIG.3. (a,b) Polarizing microscope textures of (a) the $TIC$ with no
umbilics and (b) periodic fingers pattern in a homeotropic cell with one of
the substrates rubbed along the black bar. (c,d) Light transmission through
the cell with rubbed homeotropic substrates placed between crossed
polarizers for (c) $\rho =0$ and (d) $\rho =0.5$. The insets in (c,d) show
details of intensity changes in the vicinity of homeotropic-$TIC$
transition; note that the rubbing-induced pretilt makes these dependencies
not as sharp as normally observed in non-rubbed homeotropic cells (see, for
example, Ref. \cite{Rosenblatt}).

FIG.4. Phase diagram of structures in the $U$ vs. $\rho $ parameter space
for CLCs in the cells with homeotropic boundary conditions and substrates
rubbed in anti-parallel directions. The cell has mylar spacers at the edges;
no spacer particles are present in the bulk. The lines $V0-V3$, $V01$, $V02$
separate different phases and the two dashed vertical lines mark $\rho
_{triple}=0.816$ and $\rho _{tricritical}=0.861$ corresponding to the triple
and tricritical points, similar to Fig.1. The solid line $V0$ was calculated
using Eq.(1) and is the same as in Fig.1; the solid lines $V1-V3$, $V01$, $%
V02$ connect the experimental points to guide the eye.

FIG.5. Polarizing microscope textures illustrating the transition from (a)
homeotropic untwisted state to (b) $TIC$ with no umbilics and total twist $%
\Delta \phi \approx \pi $ between the substrates, and then to (c) fingers
pattern that slowly ($\sim 1s$) appears from $TIC$, and then to (d) uniform $%
TIC$ with $\ \lessapprox 2\pi $ twist. The applied voltages are indicated.
The homeotropic cell has substrates rubbed in anti-parallel directions; $%
\rho =0.65$. The cell was assembled by using mylar spacers at the cell
edges; no particles or other nucleation sites are present in the working
area of the cell.

FIG.6. Influence of spherical particles with perpendicular surface anchoring
on the CLC structures in homeotropic cells: (a) particle-induced director
distortions in the homeotropic state; (b) director distortions in the $TIC$
at $U\approx U_{th}$; (c) $TIC$ $10ms$ after $U>U_{th}$ is applied and (d)
relaxation of the distortions in $TIC$ via formation of fingers as observed $%
\approx 1s$ after $U$ is applied (the particles become nucleation sites for
the fingers).

FIG.7. Phase diagram of structures in the $U$ vs. $\rho $ parameter space
for CLCs in homeotropic cells with rubbed substrates : (a) anti-parallel
(i.e., at 180$^{\circ }$); (b) at 90$^{\circ }$; (c) at 270$^{\circ }$. The
frequency of the applied voltage is $1kHz$, which is amplitude-modulated
with a $50Hz$ sinusoidal signal. The boundary lines $V0-V3$, $V01$, $V02$
separate different phases and the two dashed vertical lines mark $\rho
_{triple}=0.816$ and $\rho _{tricritical}=0.861$, similar to Figs.1 and 4.
The solid line $V0$ was calculated using Eq.(1) and is the same as in
Figs.1,4; the solid lines $V1-V3$, $V01$, $V02$ connect the experimental
points to guide the eye.

FIG.8. FCPM\ cross-sections (a,b) and schematic of director structures (c,d)
of $TIC$ with twist: (a,c) $\approx \pi $ at $U=5Vrms$ and $\rho =1/2$;
(b,d) $\approx 2\pi $ at $U=5Vrms$ and $\rho =1$. The polarization of the
probe light in FCPM marked by ''P'' is in the $y$-direction, along the
normal to the pictures in (a,b).

FIG.9. FCPM\ vertical cross-sections (a) and schematic of director
structures (b) of a $CF1$-type isolated finger. The polarization of the
probe light in FCPM marked by ''P'' is in the $y$-direction, along the
normal to the picture in (a). The non-singular $\lambda $-disclinations are
marked by circles in (b); the open circles correspond to the $\lambda
^{-1/2} $ and the solid circles correspond to\ the $\lambda ^{+1/2}$
disclinations.

FIG.10. FCPM\ cross-sections (a) and schematic of director structures (b) of
a periodic finger pattern composed of $CF1$s separated by homeotropic
stripes. The polarization of the FCPM probe light marked by ''P'' is normal
to the picture in (a).

FIG.11. FCPM vertical cross-section illustrating the voltage-induced
transition from (a) isolated fingers coexisting with homeotropic state to
(f) periodically-modulated $TIC$ and then to (g) a uniform $TIC$. The
fingers gradually widen (b-e) and then merge in order to form the modulated $%
TIC$ (f). The polarization of the FCPM probe light marked by ''P'' is normal
to the pictures in (a). The applied voltages are indicated, the confinement
ratio is $\rho =0.9$.

FIG.12. FCPM cross-section (a) and reconstructed director structure (b)
illustrating the $CF$ expanding into $TIC$. The polarization of the FCPM
probe light marked by ''P'' is along $y$, normal to the picture in (a).

FIG.13. FCPM cross-section (a) and reconstructed director structure (b) of $%
CF2$ finger; the $CF2$ can expand in one (c) or two in-plane directions (d)
forming $TIC$. The non-singular $\lambda $-disclinations are marked by
circles in (b); the open circles correspond to the $\lambda ^{-1/2}$
disclinations and the solid circle corresponds to\ the $\lambda ^{+1}$
disclination. The FCPM polarization is normal to the picture in (a).

FIG.14. FCPM cross-section (a) and reconstructed director structures (b,c)\
formed between the parts of a cell with different in-plane twist and helical
axis along the $z$: (b) $TIC$ with $\approx \pi $ twist, coexisting with
homeotropic untwisted state and separated by $CF2$-like structure with two
nonsingular $\lambda ^{-1/2}$ disclinations; (c) $TIC$ with $\approx 2\pi $
twist coexisting with $TIC$ with $\approx \pi \,\ $twist. The FCPM
polarization marked by ''P'' is normal to the picture in (a).

FIG.15. FCPM images and director structure of a $T$-junction of $CF1$ and $%
CF2$: (a) in-plane FCPM\ cross-section; (b) perspective view of the 3-D
director field of the $T$-junction; (c,d,e) FCPM cross-sections along the
lines $c-c$, $d-d$, and $e-e$ in part (a). The FCPM\ cross-sections in (c,e)
correspond to $CF2$ and cross-section (d) corresponds to $CF1$. The FCPM
polarization marked by ''P'' is normal to the pictures in (c,d,e).

FIG.16. FCPM cross-section (a) and reconstructed director structures (b) of $%
CF3$ finger. The FCPM polarization marked by ''P'' is normal to the picture
in (a). Two $CF3$s can be seen in (a); the director structure of only one $%
CF3$ is shown in (b). The singular disclinations at the two substrates are
marked by circles in (b).

FIG.17. FCPM\ cross-sections (a,c) and reconstructed director structure (b)
of the $CF4$ fingers. The FCPM polarization is along $y$ in (a,c). The
singular disclinations at one of the substrates are marked by circles in
(b). The $CF4$s in (c) have singular disclinations at the same substrate.

FIG.18. A representative $TIC$ equilibrium configuration obtained as
numerical solution of Eqs.~(\ref{eqn:ELa}--\ref{eqn:ELc})): (a)~tilt angle $%
\theta $, (b)~twist angle $\phi $, (c)~electric potential $\psi $. The
material parameters used in the calculations were taken for the AMLC-0010\
host doped with ZLI-811, $d=5\,\mu m$, $\rho =0.5$, $U=3.5Vrms$.


\begin{references}
\bibitem{error} We note that there is a mistake in Eq.(29) of Ref.\cite%
{Oswald-review00} as it is different from Eq.(14) of Ref.\cite%
{CHnegativeDeltaE}.
\end{references}
\end{document}